\documentclass[12pt]{article}%

\nonstopmode

\usepackage{natbib}
\usepackage{color}
\usepackage{amsmath}
\usepackage{amsbsy}
\usepackage{amscd}
\usepackage{amsopn}
\usepackage{amstext}
\usepackage{amsxtra}
\usepackage{amssymb}
\usepackage{amsthm,amsfonts}
\usepackage{graphicx}
\usepackage{epstopdf}
\usepackage{hyperref,url}
\usepackage{epsfig}
\usepackage[all]{xy}

\usepackage[T1]{fontenc}
\fontencoding{T1}  
\usepackage[utf8]{inputenc}

\newtheorem{theorem}{Theorem}

\newtheorem{definition}{Definition}

\def\beq{\begin{eqnarray}}
\def\eeq{\end{eqnarray}}
\def\beqq{\begin{eqnarray*}}
\def\eeqq{\end{eqnarray*}}
\def\beeq{\begin{eqnarray*}}
\def\eeeq{\end{eqnarray*}}
\def\be{\begin{equation}}
\def\ee{\end{equation}}
\newcommand{\bea}{\begin{eqnarray}}
\newcommand{\eea}{\end{eqnarray}}
\newcommand{\beaa}{\begin{eqnarray*}}
\newcommand{\eeaa}{\end{eqnarray*}}
\newcommand{\barr}{\begin{array}}
\newcommand{\earr}{\end{array}}

\newcommand{\benum}{\begin{enumerate}}
\newcommand{\eenum}{\end{enumerate}}

\def\BF={{\mathbb{F}}}
\def\BG={{\mathbb{G}}}
\def\BH={{\mathbb{H}}}
\def\bOne={{\bf 1}}

\def\CB{{\cal B}}
\def\CC{{\cal C}}

\def\CL{{\cal L}}

\def\CS{{\cal S}}

\def\qed{\hfill$\sqcap\kern-7.0pt\hbox{$\sqcup$}$\\}

\def\BBR{\mathbb{R}}

\def\One{{\bf 1}}

\def\ED{{\rm D}}
\def\IA{{\rm Z}}
\def\ID{{\rm X}}
\def\EE{{\rm E}}

\begin{document}

\title{
Bank Panics and Fire Sales, Insolvency and Illiquidity}
\author{T. R.  Hurd\\
\emph{ Department of Mathematics and Statistics, McMaster University, Canada }\\
}
%
\maketitle

\abstract{Banking system crises are complex events that in a short span of time can inflict extensive damage to banks themselves and to the external economy. The crisis literature has so far identified a  number of distinct effects or channels that can propagate distress contagiously both directly within the banking network itself and indirectly, between the network and the external economy. These contagious effects, and the potential events that trigger these effects, can explain most aspects of past crises, and are thought to be likely to dominate future financial crises. Since the current international financial regulatory regime based on the Basel III Accord does a good job of ensuring that banks are resilient to such contagion effects taken one at a time, systemic risk theorists increasingly understand that future crises are likely to be dominated by the spillovers between distinct contagion channels. The present paper aims to provide a model for systemic risk that is comprehensive enough to include the important contagion channels identified in the literature. In such a model one can hope to understand the dangerous spillover effects that are expected to dominate future crises. To rein in the number and complexity of the modelling assumptions, two requirements are imposed, neither of which is yet well-known or established in the main stream of systemic risk research. The first, called stock-flow consistency, demands that the financial system follows a rigorous set of rules based on accounting principles. The second requirement, called Asset-Liability symmetry, implies that every proposed contagion channel has a dual channel obtained by interchanging assets and liabilities, and that these dual channel pairs have a symmetric mathematical representation. }
\bigskip

\noindent
{\bf Key words:\ }
Systemic risk, banking network, stock flow consistency, asset liability symmetry,  cascade, interbank exposure, funding liquidity,  insolvency, indirect contagion, asset fire sales, bank panics.


\section{Introduction}
\label{sec:1}

Hyman Minsky's ``financial instability hypothesis'' \citep{Minsky82} captures the idea that of all economic sectors, the financial sector is most prone to systemic risk. As \citep{kindleberger2011manias} has investigated, prolonged periods of financial stability create the conditions, including greed and avarice, for the ``Minsky moment'' when financial instability suddenly appears. There is broad agreement \citep{KaufScot03,Taylor2010} that a Minsky moment, or financial crisis, must involve at least three attributes: (i) a triggering event, which may be narrow (affecting one bank), or broad (affecting many banks or a whole market); (ii) the propagation of shocks within the financial system;  (iii) severe impact on the macroeconomy. Major financial crises of the past typically exhibit all of the following effects: wide-scale loss of confidence, bank panics,  market crashes, a credit crunch and asset fire sales. 

Systemic risk literature reviewed in   \citep{Kaufman94,BandHart00,BandHartPeyd10,GalaMoes13} describes and develops the theoretical and empirical basis for a rich variety of ``channels'' or mechanisms that are important drivers of financial crises of the past.  These studies view the system as an interconnected network of financial institutions (``banks'') that have stylized balance sheets such as the one shown in Table \ref{basicbalancesheet2}. The studies reveal at least four fundamentally distinct channels for the propagation and amplification of shocks within the financial system and to the macroeconomy: (i) solvency contagion, caused by banks with weak equity; (ii) funding liquidity contagion, caused by banks with weak cash positions; (iii) asset fire sales; and (iv) bank runs and panics. A fully developed banking crisis  can be  expected to exhibit feedback and spillover effects between all four of these channels. 

\begin{table}[!htbp]
  \centering 
{\centering\begin{tabular}{|c||c|}
\hline
 Assets  &  Liabilities  \\ \hline\hline interbank assets $\bar\IA$
 & interbank debt $\bar\ID$ \\ \hline external fixed assets $\bar{\rm A}$& external debt $\bar\ED$  \\ \hline  external liquid assets $\bar{\rm C}$ & equity $\bar\EE$  \\
\hline
\end{tabular}}
\caption{A stylized bank balance sheet. }\label{basicbalancesheet2}
\end{table}

Systemic risk models predating the 2007-08 financial crisis focus on solvency or default contagion, as reviewed in \citep{Upper11}. Such models focus on the build up of shocks to creditor banks when one or more banks default: the review concludes that none of the 15 models studied was able to anticipate the imminent financial crisis of 2007-08. Funding liquidity contagion refers to the shocks on debtor banks that follow when their creditor banks withdraw short term funding. As studied by \citep{BrunPede09,GaiHalKap11}, this channel leads directly to credit freezes and liquidity hoarding. The third channel, called asset fire sales or market illiquidity, was studied for example in \citep{ShleVish92,CifuFerrShin05,GreLanThe15}. Large scale market effects arise when distressed banks shrink their balance sheets by selling illiquid assets. The resultant asset price impact creates mark-to-market losses to other banks holding overlapping asset portfolios, leading to contagious amplification of the crisis.  Fourthly, bank runs and panics, studied for example in \citep{CaloGort91,RochVive04}, occur when banks' depositors and wholesale funders decide to withdraw funds or equivalently, fail to roll over funding. 

In this list of four important channels of contagion, one should distinguish the so-called ``direct'' contagion channels (i) and (ii), where shocks are transmitted within the financial system, from the  ``indirect'' contagion of (iii) and (iv) which both involve feedback between the financial system and the macroeconomy (i.e. the exterior of the system). The indirect contagion channels (iii) and (iv) can also be interpreted as the drivers of the so-called ``financial accelerator'' reviewed in \citep{BernGertGilc99} that links the distress in the financial system to shocks in the macroeconomy. 

A number of recent papers build on earlier systemic risk literature by exploring the various linkages and spillovers between these contagion channels. Their overall aim is to show that important crisis amplification effects emerge from these linkages that cannot be understood when restricting attention only one or two channels.  \citep{BooPadTiv14} introduces an agent-based model for analyzing the vulnerability of the financial system to asset- and funding-based fire sales. A recent paper \citep{AldaFaia16} provides a  framework that combines the two indirect contagion channels within a financial network model.  \citep{ContScha17}  presents a framework for quantifying the impact of fire sales in a network of financial institutions holding common assets and subject to leverage or capital constraints. The book \citep{gatti2011macroeconomics} and references therein provide a general agent-based modelling framework for macroeconomics.  \cite{HusHalKokPerKra17}  present a multi-layered network model to assess the systemic implications of adopting bail-in bank resolution among the largest euro area banking groups.

The present paper will augment this recent stream of literature by introducing a sequence of extended models of systemic risk based on the framework provided by \citep{EiseNoe01}. The last and most complete of these new models unifies all four of the above contagion channels and the potential spillovers between them. Our primary purpose is illustrative, and therefore we strive in each model for the simplest possible explanation for each channel, and consider only the most direct linkages between channels.  Despite its known deficiencies as a picture of systemic risk, the Eisenberg-Noe 2001 model is simple and well-studied. For this reason we adopt it as the conceptual basis for our new framework, and like that model, treat the financial system as a network of interbank linkages between $N$ banks, coupled to the macroeconomy (system exterior). The banks follow deterministic behavioural rules of the type developed in the systemic risk literature reviewed in  \citep{Upper11}. This places our framework close to the literature on agent-based models in economics \citep{FarmerFoley09}. The triggering event at the onset of the financial crisis being modelled will put the system in a severely stressed state with some banks either insolvent or illiquid. Thereafter, the contagion dynamics proceeds stepwise in time until eventually an equilibrium is approached that represents the end of the crisis. The following requirements underlie all the models proposed in this paper. They are made for specificity and to rein in complexity, but can be replaced later by more complex, realistic assumptions.
\begin{enumerate}
  \item[A1] The time step is one day.  Prior to the triggering event on day $n=0$, all banks are in their normal, solvent and liquid state. The triggering event causes some banks to become either insolvent or illiquid. 
  \item[A2] Bank balance sheets have the structure shown in Table \ref{basicbalancesheet2} and are recomputed daily on a mark-to-market, stock-flow consistent basis.
 \item[A3]  Interest payments and dividends are neglected during the crisis.
  \item[A4] The dynamics of balance sheets exhibit a formal symmetry between assets and liabilities.
\end{enumerate}

The stock-flow consistency  assumption A2  requires the balance sheet dynamics to be decomposable into a sequence of elementary steps which incorporate so-called {\em quadruple entry accounting} as described in the book  \citep{GodlLavo12}. This imposes a system-wide logical structure to the valuation of assets and liabilities. It ensures that anyone can verify that the proposed model dynamics, no matter how complex, avoids any basic economic pitfall. In subsequent discussions, we will find it helpful to propose the existence of an imaginary {\em system regulator or auditor} that forces all banks to follow stock-flow consistent rules based on market-to-market valuation of balance sheet entries.

The proposed asset-liability (AL) symmetry invoked in A4 extends the natural left-right symmetry shown in Table 1 to the level of the dynamical changes of the system. Thus, for example, the sale of \$1 worth of fixed asset leads to the balance sheet adjustments: $\Delta\bar{\rm A}=-1, \Delta\bar{\rm C}=1, \Delta\bar{\rm D}=\Delta\bar{\rm E}=0$. This is AL symmetric to a mark-to-market drop in the value of the external debt: $\Delta\bar{\rm D}=-1, \Delta\bar{\rm E}=1, \Delta\bar{\rm A}=\Delta\bar{\rm C}=0$. Invoking AL symmetry exemplifies the Principle of Occam's Razor,  the criterion that values explanations when they are based on the fewest (or simplest) possible modelling assumptions.  Asking for AL symmetry in systemic risk models is motivated by an observation made by \citep{Hurd16a} and others that existing models of solvency contagion are formally symmetric to models of funding liquidity contagion under the interchange of debtor and creditor relationships.  While current and prospective accounting principles do not treat assets and liabilities symmetrically, nonetheless, commentary on the IFRS new accounting regulatory measures being implemented by the International Accounting Standards Board (IASB) suggests that future revisions will move in the direction of making financial regulation and definitions more exactly symmetric between assets and liabilities\footnote{From an IASB website\\
{\tt www.iasplus.com/en/meeting-notes/iasb/2013/iasb-february-2013/cf-recognition-derecognition}
\\ 
 ``... Some Board members were concerned with the drafting of the language specifically that it was very definite that an asset would always exist for one party where another has a liability. Board members were in tentative agreement over the requirement for symmetry in the DP (discussion paper) but it was still unclear as to whether there would be absolute symmetry (and how symmetry would affect the Framework) and whether if a liability exists for one party, an asset will always exist for the other party.''
 }. Since AL symmetry is strong at the level of financial securities,  it is appropriate to explore the implications of a weak type of AL symmetry to simplify the development of unified solvency and liquidity cascade models.  We emphasize that this symmetry is ``weak'' in the sense that while the structure of the dynamical equations is AL symmetric,  the actual parameter values are not symmetric. Taking this abstract AL symmetry seriously leads to certain insights. For example, it may seem surprising that  ``equity'', defined as assets minus debt, is a liability that  can be seen as the dual concept to ``cash''. 

The problem of solvency cascades in interbank networks has been studied for over two decades. The Eisenberg-Noe 2001 model stands at the centre of such research as the fundamental characterization of solvency cascades in a network of ``banks'' with stylized balance sheets. Subsequent research into solvency cascades, notably works by \citep{ElsLehSum06,ElsLehSum06a,NieYanYorAle07}, and most of the studies reviewed in \citep{Upper11} reference this model. A more recent paper, \citep{GaiKapa10a}, offers an even simpler model of solvency cascades that ties this subject into the general theory of information cascades on  random networks stemming from the famous model of \citep{Watts02}.

More recently, the problem of funding illiquidity has been recognized and studied as an additional channel for contagion in financial networks.  A model introduced in \citep{GaiHalKap11} provides us with a picture that banks, when hit by an external shock to their liabilities, such as a large withdrawal from a depositor, may experience ``stress'', or an insufficiency in liquid assets. Such a bank is then assumed to react by ``hoarding'' liquidity, that is recalling short term loans from debtors, including other banks. This creates the mechanism for liquidity contagion: recalling interbank loans transmits liquidity shocks to debtor banks. Interestingly, the mathematical equations for the funding liquidity cascade models of GHK 2011 and \citep{SHLee13} are identical to variations of the EN solvency cascade model, but with the direction of shocks reversed. This is a basic consequence of AL symmetry in systemic risk.  A brief review of these two models and their generalizations, touching on their analogy to the EN 2001 model, can be found in Section 2.2 of  \citep{Hurd16a}. 

The present paper achieves four main results:
\begin{enumerate}
  \item Theorem \ref{SFC} proves that the EN 2001 model has an equivalent solvency cascade formulation satisfying A1, A2 and A3 of the four modelling requirements. A natural interpretation of the EN 2001 cascade invokes a {\em system regulator} that provides an orderly resolution of insolvent banks through a bail-in mechanism applied iteratively.  
  \item The generalized liquidity (GL) cascade model  is introduced and shown to be dual to the EN model under the AL symmetry, and therefore it also satisfies A1, A2 and A3. The interpretation is that the {\em system regulator}  functions as the ``lender of last resort'' (LLR)  to resolve illiquid banks by providing overdrafts with strict repayment rules. 
  \item The above solvency and liquidity cascade models are unified symmetrically and  unambiguously as a complete model of {\em direct contagion}, called the Solvency and Liquidity cascade model (SL cascade model). This unification now satisfies all of assumptions A1 to A4. Theorem \ref{SLtheorem} provides conditions that imply the state of the system must converge monotonically to a cascade equilibrium.   
\item  The channels of indirect contagion, asset fire sales and bank runs and panics, are integrated with the SL contagion model, yielding the Extended  Solvency and Liquidity cascade model (ESL cascade model) that unifies in a symmetric way the four most important contagion channels discussed in the literature. Figure \ref{crisiscycle} provides a schematic representation of this model. 
\end{enumerate}

The remainder of this paper is structured as follows. In Section 2, we review the EN solvency cascade model, extended to allow external debt to be greater than external assets, and prove Theorem \ref{SFC} which shows that the model has a stock-flow consistent representation. In this formulation of the EN model, there are two levels of bank insolvency.  ``Partial insolvency'' is when a bank is able to repay external depositors in full, but only able to partially repay their interbank debt. ``Full insolvency'' is when the bank is unable to repay any interbank debt. Next, the GL cascade model is defined and shown to be an exact dual of the EN model, with cash assets behaving as the dual counterpart to equity. In this model, a bank that has overdrawn its cash asset is called ``partly illiquid'' if it can realize sufficient cash by recalling their interbank loans, and ``fully illiquid'' if it needs to also liquidate fixed assets to fulfil its cash obligations. 
In Section 3, we formulate and interpret the SL cascade model, a stock-flow consistent model that unifies the solvency and liquidity cascades from Section 2. The two channels of indirect contagion, asset fire sales and bank panics, are then integrated with the SL cascade model  in Section 4, leading to the Extended SL (ESL) cascade model. 
The concluding Section 5 offers some points of view on further questions about systemic risk that are amenable to study within the network framework developed in this paper. 

\section{Financial Network Cascade Models}
\label{sec:2}

This section will review systemic risk models that view a crisis as a contagious cascade on a network of $N$ ``banks'', labelled by $i=1,2,\dots,N$ (which may include non-regulated leveraged financial institutions such as hedge funds, or other regulated financial institutions such as insurance companies). These institutions are linked by a collection of exposures such as interbank loans or derivative exposures. Table \ref{basicbalancesheet2} shows a schematic characterization of their balance sheets.

Prior to the onset of the crisis, a bank's balance sheet $\bar {\rm B}=[\bar \IA,\bar{\rm A}, \bar{\rm C},\bar \ID,\bar{\rm D}, \bar{\rm E}]$ (labelled by barred quantities)  consists of {\em nominal values} \index{nominal value} of assets and liabilities, which correspond to the aggregated values of the contracts, valued as if all banks are solvent. Nominal values can also be considered {\em book values} \index{book values} or {\em face values}\index{face values}. Assets (loans) and liabilities (debts) are decomposed into interbank and external quantities depending on whether the counterparty is a bank or not. Banks and institutions that are not part of the system under analysis are deemed to be part of the exterior, and their exposures are included as part of the external debts and assets. Finally, two categories of external assets are included. Fixed assets model the retail loan book and realize only a fraction of their value if liquidated prematurely while liquid assets includes government treasury bills and the like that are assumed to be as liquid as cash.

\begin{definition}\label{balancesheet} The {\em nominal value of assets} of bank $i$ prior to the crisis consists of the {\em nominal interbank assets}  ${\bar \IA}_i$, the {\em nominal external fixed assets} $\bar{\rm A}_i$, and the {\em nominal external liquid assets} $\bar{\rm C}_i$. The {\em nominal value of liabilities} of the bank consists of the {\em nominal interbank debt} ${\bar \ID}_i$, the {\em nominal external debt} ${\bar \ED}_i$ and the bank's {\em nominal equity}  ${\bar \EE}_i$. The {\em nominal exposure} of bank $j$ to bank $i$ is denoted by $\bar\Omega_{ij}$. All components of $\bar{\rm B}$ and $\bar \Omega$ are non-negative, and the {\em accounting identities} are satisfied:
\beq
\nonumber &&{\bar \IA}_i=\sum_j \bar \Omega_{ji}, \quad {\bar \ID}_i=\sum_j\bar \Omega_{ij},\quad \sum_i {\bar \IA}_i=\sum_i {\bar \ID}_i,\quad \bar \Omega_{ii}=0\ ,\\
\label{accounting}
&&{\bar \IA}_i +\bar{\rm A}_i+\bar{\rm C}_i={\bar \ID}_i+{\bar \ED}_i+{\bar \EE}_i\ . \eeq
\end{definition}


As discussed in detail in Chapter 2 of the book \citep{GodlLavo12}, stock-flow consistent modelling requires using {\em actual} or {\em mark-to-market} values during the cascade, as balance sheets move away from their nominal values. We will denote these  mark-to-market values  by symbols ${\rm B}=[ \IA,{\rm A}, {\rm C}, \ID,{\rm D},{\rm E}]$ and $\Omega$ without upper bars.
Equity $\EE$ represents the total firm value held by shareholders. One school of thought defines equity as the market capitalization of the firm, and an additional balance sheet entry called {\it net worth} is needed to enforce the balancing of assets and liabilities. Instead, we define $\EE$ so that the accounting identity \eqref{accounting} holds for both nominal values and actual values at all times. Equivalently, we assume the net worth is zero at all times. Equity can be regarded as a buffer that, while positive, absorbs market shocks. Economic cascade models invoke the notion of {\it limited liability}, and define a firm to be {\em defaulted} or {\em insolvent} (for us, these two terms mean the same) when its equity is non-positive. By the accounting identity, this means its aggregated assets are insufficient to pay the aggregated debt, and such a bank is assumed to be unable continue to operate as a going concern. In a somewhat analogous way, liquid assets ${\rm C}$ are kept available to pay deposit withdrawals on demand. We define a bank to be {\em illiquid} when ${\rm C}$ is non-positive: such a bank may have difficulties in raising enough cash to become compliant with the accounting constraints.

\begin{definition} \begin{enumerate}
  \item A {\em defaulted bank} or {\em insolvent bank} is a bank with $\EE\le 0$. A {\em solvent bank} is a bank with $\EE>0$.
 \item An {\em illiquid bank} is a bank with ${\rm C}\le 0$. A {\em liquid bank} is a bank with ${\rm C}>0$.
 \item A bank that is either illiquid or insolvent is called {\em impaired}. 
\end{enumerate}\end{definition}

A bank whose equity is dangerously low, or negative, has few strategies available to restore financial health. Raising additional equity through the issue of new shares dilutes existing shareholders, because prospective buyers are aware that their investment is adding value to the bank's debt. The alternative of restructuring debt through negotiation also sends a very unsavory signal to the market. The most likely response of such a weak bank will be to sell some assets to reduce its leverage. Similarly, a bank whose liquid assets have been depleted through a run of deposit withdrawals will be challenged to find new depositors elsewhere and will lose equity if illiquid assets are sold at a loss. Because such stressed banks have limited options available, it is reasonable to create models in which their behaviour is always determined by the current state of their balance sheets. We will call this programmed behaviour by stressed banks the {\em cascade mechanism}. 

We will suppose that our schematic economic crisis is triggered by a major macroeconomic event. This trigger event causes a large negative shock to the collection of bank balance sheets, putting the system into a {\it stressed} initial state with balance sheets $ {\rm B}^{(0)}_i=\bar {\rm B}_i+\delta  {\rm B}_i$. A typical trigger shock will involve one or both of the following elementary balance sheet changes to one or more banks:
\begin{enumerate}
  \item An asset price shock causes changes $\delta{\rm A}=\delta{\rm  E}<0$. 
  \item A deposit withdrawal shock causes changes $\delta{\rm D}=\delta{\rm  C}<0$. Note that such a demand withdrawal must be repaid immediately in cash in full. 
\end{enumerate}

 The static cascade models we shall discuss in this section all share the timing and operational assumptions A1-A3 listed in Section 1, which imply  a {\em cascade process} that proceeds in daily steps.  The heuristic picture to have in mind is that the cascade is a systemic crisis that develops quickly. As it proceeds a number of banks become impaired, either defaulted or illiquid, and other banks adopt defensive behaviour to protect themselves from the contagion. In all our examples, the resultant cascade process leads to a sequence of collections of daily balance sheets $\CB^{(n)}:=[{\rm B}^{(n)}_i]_{i\in[N]}$ and exposures $\Omega^{(n)}:=[\Omega^{(n)}_{ij}]_{i,j\in[N]}$ for $n\ge 0$. Each bank's balance sheet has the form ${\rm B}^{(n)}_i:=[\IA^{(n)}_i, {\rm A}^{(n)}_i, {\rm C}^{(n)}_i,\ID^{(n)}_i,\ED^{(n)}_i,{\rm E}^{(n)}_i]$. The accounting identity \eqref{accounting} is assumed to hold at all times. However, some banks will be impaired, meaning either ${\rm C}^{(n)}_i<0$ or ${\rm E}^{(n)}_i<0$, and our models must specify the protective strategies impaired banks will adopt. We shall call these behavioural assumptions the {\em cascade mechanism}, and need to explore the impact of this behaviour on both the financial system and its exterior.    

\medskip\noindent{\bf Notation:\ } We adopt a vector notation with banks indexed by the set $[N]:=\{1,\dots,N\}$.  We define ${\bf 0}=[0,\dots, 0],\ 
{\bf 1}=[1,\dots, 1]\in\BBR^N$ and for vectors $x=[x_i]_{i\in [N]},y=[y_i]_{i\in [N]}\in\mathbb{R}^N$  relations
\beqq
x \le y &\mbox{means}&  \forall\ i,\ x_i\le y_i\ ,\\
x < y &\mbox{means}& x\le y,\ \exists\ i: x_i< y_i\ ,  \\
\min(x,y)&=&[\min(x_i,y_i)]_{i\in [N]}\ ,\\
\max(x,y)&=&[\max(x_i,y_i)]_{i\in [N]}\ ,\\
(x)^+&=&\max(x,{\bf 0})\,\\
(x)^-&=&\max(-x,{\bf 0})\ =\ -\min(x,{\bf 0})\ .
\eeqq
Whenever $x\le y$ we define the hyperinterval $[x,y]=\{z:x\le z\le y\}$. 
The {\em indicator function} of a condition $P$ has the values
\[ \One (P)=\left\{\begin{array}{ll }
   1   &  \mbox{ if $P$ holds,} \\
    0  &   \mbox{ otherwise.}
\end{array}\right.\]

\subsection{The Eisenberg-Noe Model}
The  paper \citep{EiseNoe01} addressed the problem of finding the consistent amounts that bank clients should pay in a centrally cleared payment system when one or more client banks are insolvent and unable to pay their full obligation. As many other papers have noted, a slight extension of their framework also serves as a very basic model that determines the equilibrium collection of mark-to-market balance sheets that represents the end result of a systemic crisis triggered by a shock that causes one or more bank insolvency. Following \citep{Hurd16a}, we review this extended model, henceforth called the EN model, and show how it can be reinterpreted as an example of a static solvency cascade model satisfying all of Assumptions A.  

We assume that the post-trigger  bank balance sheets and exposures $\CB^{(0)}, \Omega^{(0)}$ are as shown in Table \ref{basicbalancesheet2} and in addition 
 \begin{enumerate}
  \item[EN1] External debt is senior to interbank debt, and all interbank debt is of equal seniority; 
  \item[EN2] There are no losses due to bankruptcy charges.
\end{enumerate} 

\noindent{\bf Remark:\ } Unlike \citep{EiseNoe01}, we do not assume ${\rm A}^{(0)}_i+{\rm C}^{(0)}_i-\ED^{(0)}_i\ge 0$. Moreover, since liquidity is not a concern in the EN model, we need not distinguish ${\rm A}^{(0)}$ from ${\rm C}^{(0)}$.

We focus on the {\em fractional clearing vectors} ${\rm p}=[p_1,\dots, p_N]$ representing the fractional recovery values of defaulted interbank debt, in the post-crisis equilibrium. Again, this differs slightly from  \citep{EiseNoe01}: in terms of fractional clearing values, their {\em clearing vectors} are $[p_1\ID^{(0)}_1,\dots, p_N\ID^{(0)}_N]$, with the convention that $p_i=0$ for any bank with $\ID^{(0)}_i=0$. The collection of {\em fractional clearing vectors} ${\rm p}$ representing the end of the crisis is identified with the collection of fixed points of the equation ${\rm p}={\rm F}({\rm p})$ where ${\rm F}=[F_1,\dots, F_N]$ with
 \be
 \label{ENmapping}  F_i({\rm p}):=\One(\ID^{(0)}_i>0)  \min\Bigl(1,\max\bigl(0,({\rm A}_i^{(0)}+{\rm C}_i^{(0)} -{\ED}^{(0)}_i+\sum_j\Omega^{(0)}_{ji}p_j)/\ID^{(0)}_i\bigr)\Bigr)  \ .\ee
This equation has the meaning that the amount of interbank debt bank $i$ can pay depends on the interbank asset values realized from all other banks.  A bank with $p_i=1$ is {\em solvent} and  can pay its full interbank debt; a bank with a fractional recovery $0<p_i<1$ is called {\em partially insolvent}; a {\em fully insolvent} bank has $p_i=0$ and cannot pay all of its external debt. By similar reasoning, the fractional recovery of the external debt of bank $i$ is given by
\be q_i=\One({\ED}^{(0)}_i>0)\ \min\bigl(1,({\rm A}_i^{(0)}+{\rm C}_i^{(0)}+\sum_j\Omega^{(0)}_{ji}p_j )/{\ED}^{(0)}_i\bigr)  \ .
\ee

The main theorem of \citep{EiseNoe01}, slightly generalized, is 
\begin{theorem} \label{ENThm1} Corresponding to every post-trigger financial system $\CB^{(0)},\Omega^{(0)}$ satisfying Assumptions EN1, EN2 there exists a greatest and a least fractional clearing vector $\hat{\rm p}$ and $\check{\rm p}$. 
 \end{theorem} 

The proof of this theorem found in  \citep{EiseNoe01} is based on the Knaster-Tarski Fixed Point Theorem which states that the set of fixed points of a monotone mapping of a complete lattice onto itself, is a complete lattice. In the above context, the mapping ${\rm F}$ is monotone increasing from the hyperinterval (a complete lattice) $[{\bf 0},{\bf 1}]$ onto itself. Since every complete lattice is non-empty and has a maximum and minimum, the theorem follows.

\subsubsection{Stock-flow Consistency}
A stronger result than Theorem \ref{ENThm1} is also true because the mapping  ${\rm F}$ is everywhere continuous. Continuity implies that  $\hat{\rm p}$ must be the limit of the decreasing sequence of iterates ${\rm p}^{(n)}:={\rm F}({\rm p}^{(n-1)})$  starting from ${\rm p}^{(0)}={\bf 1}$. Similarly, $\check{\rm p}$ will be the limit of the increasing sequence of iterates starting from ${\rm p}^{(0)}={\bf 0}$. Continuity of $\rm F$ thus allows us to reinterpret the greatest fractional clearing vector $\hat{\rm p}$ as the end result of a cascade process with the daily updating of the {\em cascade mapping} ${\rm p}^{(n)}:={\rm F}({\rm p}^{(n-1)})$ that starts at time $0$ from ${\rm p}^{(0)}={\bf 1}$.  At step $n$, the balance sheets and exposures satisfy the accounting identities \eqref{accounting} and have the form ${\rm B}^{(n)}_i=[\IA_i^{(n)}, {\rm A}_i^{(0)}, {\rm C}^{(0)}_i,\ID^{(n)}_i,\ED^{(n)}_i,{\rm E}^{(n)}_i]$ with  $\ED^{(n)}_i = q^{(n)}_i \ED^{(0)}_i$ and $ \Omega^{(n)}_{ij}=p^{(n)}_i  \Omega^{(0)}_{ij}$. We now show how the full sequence of daily balance sheets $\CB^{(n)}$ and exposures $ \Omega^{(n)}$ connect in a way that obeys Assumptions A1, A2, and A3, and in particular how the cascade mapping can be broken down further into stock-flow consistent steps.

It turns out to be useful to track each bank's {\it default buffer}\index{default buffer} 
\be\label{bufferdef}
\Delta^{(n)}_i:={\rm A}_i^{(0)}+{\rm C}_i^{(0)}+\sum_jp^{(n)}_j\Omega^{(0)}_{ji}-\ED^{(0)}_i-\ID^{(0)}_i\ . \ee
Note that the difference $\sum_j(1-p^{(n)}_j)\Omega^{(0)}_{ji}$ between  $\Delta^{(0)}_i:=\EE^{(0)}_i$ and $\Delta^{(n)}_i $ is a measure of the total impact on the equity of bank $i$ after $n$ days due to losses in its interbank assets.
The vectors ${\rm p}^{(n)}$ and $\Delta^{(n)}$ are  expressible recursively in terms of ${\rm p}^{(n-1)}$ and  $\Delta^{(n-1)}$ through the equations 
\beq
   p^{(n)}_i    &=&\One(\ID^{(0)}_i>0) \ h(\Delta^{(n-1)}_i/ \ID^{(0)}_i) \label{ENcascadeP}\\ 
   \Delta^{(n)}_i   &=&   \Delta^{(0)}_i -\sum_j\Omega^{(0)}_{ji}(1-p^{(n-1)}_j )\label{ENcascadeD}\
 \eeq
where the normalized {\em threshold function} $h$ from the extended real line $[-\infty,\infty]$ to the unit interval $[0,1]$ is
\be\label{threshold} h(x)=(x+1)^+-x^+=\min(1,\max(0,x+1)\  .\ee
The vector $ \Delta^{(n-1)}$ also determines the recovery fraction of external debt at step $n$
\be
q^{(n)}_i   =\One(\ED^{(0)}_i>0) \ h((\Delta^{(n-1)}_i+\ID^{(0)}_i)/ \ED^{(0)}_i)\ .  \label{ENcascadeQ}
\ee

 The $n$th step of the EN cascade can be given a stock-flow consistent interpretation as an attempt by the system to resolve defaulted banks simultaneously through an orderly {\it bail-in} restructuring of their debt \citep{AvgoGood15}, followed by a mark-to-market update of interbank assets.   Precisely,  we will verify that the $n$th {\it solvency step} $\CS$ given by \eqref{ENcascadeP},\eqref{ENcascadeD},\eqref{ENcascadeQ} can be broken down into three more elementary steps as follows
  \begin{enumerate}
  \item[S1] For every bank $i$ with $-\ID^{(n-1)}_i\le  {\rm E}^{(n-1)}_i<0$, all of the interbank debts of $i$ are {\it restructured}, or revalued at a fraction $\dot p^{(n)}_i=\frac{\ID^{(n-1)}_i+{\rm E}^{(n-1)}_i}{\ID^{(n-1)}_i}<1$.  If  \\${\rm E}^{(n-1)}_i<-\ID^{(n-1)}_i$, then $\dot p^{(n)}_i=0$ and the external debt of $i$ is now restructured, or revalued at a fraction $\dot q^{(n)}_i=\frac{\ED^{(n-1)}_i+\ID^{(n-1)}_i+{\rm E}^{(n-1)}_i}{\ED^{(n-1)}_i}<1$. Solvent banks are not restructured, so one has in general  \beq
\label{CS1} \dot p^{(n)}_i&=&\One( \ID_i^{(n-1)}>0)\ h({\rm E}^{(n-1)}_i/\ID_i^{(n-1)})\ ,\ \\
 \label{CS2}\dot q^{(n)}_i&=&\One( \ED_i^{(n-1)}>0)\ h\bigl((\ID_i^{(n-1)}+{\rm E}^{(n-1)}_i)/\ED_i^{(n-1)}\bigr)\ .
\eeq 
  \item[S2] All of the interbank exposures on insolvent banks $i$ are revalued, $\ID^{(n)}_i=\dot p^{(n)}_i\ID^{(n-1)}_i$, $ \Omega^{(n)}_{ij}=\dot p^{(n)}_i\Omega^{(n-1)}_{ij}$, and the external debt is revalued at $\ED^{(n)}_i=\dot q^{(n)}_i\ED^{(n-1)}_i$. Based on these revaluations, each insolvent bank $i$ now has precisely zero equity since ${\rm A}_i^{(0)}+{\rm C}_i^{(0)}+\IA^{(n-1)}_i=\ED^{(n)}_i+\ID^{(n)}_i$. 
  \item[S3] However, interbank assets must also be revalued: All banks $j$ now mark-to-market their interbank assets in light of the restructured debt of insolvent banks, meaning that $\IA^{(n)}_j=\sum_i \Omega^{(n)}_{ij}$. This revaluation leads to new equity values  \be
  \label{CS3} {\rm E}^{(n)}_i=  {\rm A}_i^{(0)}+{\rm C}_i^{(0)}+\IA^{(n)}_i -\ED_i^{(n)}-\ID_i^{(n)}=({\rm E}^{(n-1)}_i)^+-\sum_j(1-\dot p^{(n)}_j) \Omega^{(n-1)}_{ji}\ .
  \ee
\end{enumerate} 

The following theorem confirms that this defines a stock-flow consistent mapping $(\CB^{(n)},\Omega^{(n)}) =\CS(\CB^{(n-1)}, \Omega^{(n-1)})$ that generates all steps of the EN 2001 cascade. This result is important because it breaks down the cascade into elementary steps that are demonstrably consistent with stylized accounting principles. \begin{theorem}
 \label{SFC} Let the collection of initial balance sheets ${\rm B}^{(0)}_i:=[\IA^{(0)}_i, {\rm A}^{(0)}_i, {\rm C}^{(0)}_i=0,\ID^{(0)}_i,\ED^{(0)}_i,{\rm E}^{(0)}_i]$ and initial interbank exposures $ \Omega^{(0)}_{ij}$ be a financial system satisfying the accounting identities \eqref{accounting}. Let the sequence of balance sheets and exposures $(\CB^{(n)},\Omega^{(n)})$ be defined iteratively for $n>0$ by the steps {\rm S1-S3}, so in particular equations \eqref{CS1}, \eqref{CS2}, \eqref{CS3} hold. 
 Then the sequence of  recovery fractions $ p^{(n)}_i=\prod_{m=1}^{n}\dot p^{(m)}_i, q^{(n)}_i=\prod_{m=1}^{n}\dot q^{(m)}_i$ and solvency buffers defined by 
 \be
 \label{CS4} \Delta^{(n)}_i={\rm E}^{(n)}_i-\sum_{m=0}^{n-1}({\rm E}^{(m)}_i)^-
 \ee
 satisfies the EN cascade formulas \eqref{ENcascadeP},  \eqref{ENcascadeD} and \eqref{ENcascadeQ}.
\end{theorem}

Note that  equations \eqref{CS1},\eqref{CS3} give an autonomous discrete dynamics for the sequence of vector pairs ${\rm p}^{(n)},{\rm E}^{(n)}$. This dynamics depends on the values of $\ID^{(0)},{\rm E}^{(0)}$ and $ \Omega^{(0)}$ but does not depend on the values of ${\rm A}^{(0)}, {\rm C}^{(0)},\ED^{(0)}$. Impact on external debt arises only through \eqref{CS2}, which brings in dependence on $\ED^{(0)}$. Finally, external assets, both liquid and fixed, remain unchanged during the EN solvency cascade. 

\begin{proof} First, \eqref{ENcascadeD} will follow from \eqref{CS4} and $\Delta^{(0)}_i={\rm E}^{(0)}_i$ if we show 
\be\label{Diter}\Delta^{(n)}_i=\Delta^{(n-1)}_i+{\IA}^{(n)}_i-{\IA}^{(n-1)}_i, \quad  n>0\ .\ee
In the case ${\rm E}^{(n-1)}_i>0$ then certainly $({\rm E}^{(m)}_i)^{-}=0$ for all $m<n$ and hence \eqref{CS4} implies $\Delta^{(m)}_i={\rm E}^{(m)}_i$ for all $m\le n$. In this case, \eqref{CS3} implies \eqref{Diter} for all $m\le n$. In the opposite case ${\rm E}^{(n-1)}_i\le 0$, then \eqref{CS3} implies that ${\rm E}^{(n)}_i ={\IA}^{(n)}_i-{\IA}^{(n-1)}_i$. From \eqref{CS4}, because $({\rm E}^{(n-1)}_i)^-=-{\rm E}^{(n-1)}_i$, we find that again  \eqref{Diter} holds: 
\beqq
\Delta^{(n)}_i&=&{\rm E}^{(n)}_i-\sum_{m=0}^{n-1}({\rm E}^{(m)}_i)^-={\rm E}^{(n)}_i+{\rm E}^{(n-1)}_i-\sum_{m=0}^{n-2}({\rm E}^{(m)}_i)^-\\
&=&{\IA}^{(n)}_i-{\IA}^{(n-1)}_i+\Delta^{(n-1)}_i\ .
\eeqq

For $p^{(n)}_i $ as defined by the Theorem, we have ${\rm X}^{(n)}_i=p^{(n)}_i {\rm X}^{(0)}_i = \dot p^{(n)}_i {\rm X}^{(n-1)}_i $.  Now \eqref{CS1} implies that $\dot p^{(n)}_i {\rm X}^{(n-1)}_i =({\rm X}^{(n-1)}_i+{\rm E}^{(n-1)}_i)^+-({\rm E}^{(n-1)}_i)^+$. From the equation 
\[\Delta^{(n-1)}_i={\rm E}^{(n-1)}_i-({\rm X}^{(0)}_i-{\rm X}^{(n-1)}_i)-({\rm D}^{(0)}_i-{\rm D}^{(n-1)}_i)
\] and that facts that ${\rm E}^{(n-1)}_i> -{\rm X}^{(n-1)}_i$ implies ${\rm D}^{(0)}_i={\rm D}^{(n-1)}_i$ and ${\rm E}^{(n-1)}_i> 0$ implies ${\rm X}^{(0)}_i={\rm X}^{(n-1)}_i$ 
it follows that $({\rm X}^{(n-1)}_i+{\rm E}^{(n-1)}_i)^+=({\rm X}^{(0)}_i+\Delta^{(n-1)}_i)^+$  and $({\rm E}^{(n-1)}_i)^+=(\Delta^{(n-1)}_i)^+$. Therefore $p^{(n)}_i {\rm X}^{(0)}_i = ({\rm X}^{(0)}_i+\Delta^{(n-1)}_i)^+-(\Delta^{(n-1)}_i)^+$ which is equivalent to \eqref{ENcascadeP}. The proof that \eqref{ENcascadeQ} holds for $q^{(n)}_i$ as defined by the Theorem is similar. 

\end{proof}
%
%
%
%
%
%
%
%

\medskip\noindent{\bf Modifications and Extensions:\ }
There are several simple but important possible modifications that will add realism and flexibility to the above foundation solvency cascade model in future work. 
\begin{enumerate}
  \item Bankruptcy Costs:
The threshold function $h(x)$  encodes the ``zero bankruptcy costs'' assumption EN2 corresponding to a ``soft'' type of default whereas in reality we expect that bankruptcy charges will not be negligible, and may severely impact the course of a financial crisis. Losses given default can be encoded using modified threshold functions:\begin{enumerate}
  \item In the paper \citep{GaiKapa10a}  interbank debt on defaulted banks is assumed to recover zero value. Their cascade model has the EN 2001 form with a single modification that replaces the threshold function $h(x)$ that appears in \eqref{ENcascadeP}  by
\[ \tilde h(x)=\One(x\ge 0)\]
  They justify zero recovery with the statement\footnote{\citep{GaiKapa10a}, footnote 9.}: ``This assumption is likely to be realistic in the midst of a crisis: in the immediate aftermath of a default, the recovery rate and the timing of recovery will be highly uncertain and banks' funders are likely to assume the worst-case scenario.'' 
 \item For any two positive fractions $R_1,R_2<1$, a fractional loss of $(1-R_1)$ paid on interbank debt at the instant of primary default followed by a fractional loss of $(1-R_2)$ paid on external debt at the instant of secondary default can be modelled by replacing $h(x)$ by $h^{(R_1)}(x)$  in \eqref{ENcascadeP} and by $h^{(R_2)}(x)$  in \eqref{ENcascadeQ} where
 $h^{(R)}(x)=(1-R)\tilde h(x)+Rh(x/R)$.
\end{enumerate}  
\item Debt Seniority: Assumption EN1 can be replaced by more general seniority structures. For example, portions of the external debt might have seniority equal to the interbank debt. This can be accommodated by the trick of introducing a ``fictitious bank'' with label $i=0$, that never defaults and lends amounts $\bar\Omega_{i0}$ to other banks $i\in[N]$.  \citep{GourHeamMont12} studied solvency cascades where banks may have equity crossholdings and debt structured with multiple seniority tranches. 
\end{enumerate}  

\subsection{Generalized Liquidity Cascade Model}
The sum over debtor banks in the cascade mapping \eqref{ENcascadeD} demonstrates that solvency cascades are characterized by shocks that are transmitted ``downstream'', from defaulting banks to the asset side of their creditor banks' balance sheets. Two models, by \citep{GaiHalKap11} and \citep{SHLee13},  illustrate that (funding) liquidity cascades are a systemic phenomenon in which shocks are naturally transmitted ``upstream'', from creditor banks to their debtors. Next we extend and unify these two liquidity cascade models, and interpret the resultant ``Generalized Liquidity Cascade Model'' (GL cascade model), first introduced in \citep{Hurd16a} (Section 2.2.3), as a stock-flow consistent scheme. Just as a solvency cascade results when the system attempts to resolve insolvent banks, liquidity cascades will be understood to arise as the system attempts to resolve illiquid banks, that is banks with ${\rm C}_i<0$. Such illiquid banks are compelled to raise additional cash by selling or liquidating assets, either interbank assets $\rm Z$ or  fixed assets $\rm A$.

We consider again a sequence of bank balance sheets  ${\rm B}^{(n)}_i:=[\IA^{(n)}_i, {\rm A}^{(n)}_i, {\rm C}^{(n)}_i,$\\$\ID^{(n)}_i,\ED^{(n)}_i,{\rm E}^{(n)}_i]$ as shown in Table \ref{basicbalancesheet2}, and interbank exposures $ \Omega^{(n)}_{ij}$. In addition to Assumptions A, the GL cascade model also assumes
 \begin{enumerate}
  \item[GL1] Illiquid banks raise cash by first selling interbank assets. When interbank assets have been completely sold, illiquid banks then sell external assets. At each step, illiquid banks sell exactly the minimal amount of assets needed to repay their depositor withdrawals; 
  \item[GL2] There are no losses due to market price impact when interbank and external assets are sold.
\end{enumerate} 
 In this baseline version of the model, the unrealistic assumption GL2 allows us to focus solely on liquidity and to avoid discussion of solvency issues. This deficiency will be corrected in Section 4.

The event triggering the systemic crisis on day $0$ is assumed to be a vector of deposit withdrawal shocks $\delta {\rm D}=\delta {\rm C}\le 0$ that hits any or all of the banks and leaves some banks illiquid. The initial balance sheets have the form ${\rm B}^{(0)}=\bar{\rm B}-\delta{\rm B} $ with some initially illiquid banks having ${\rm C}^{(0)}=\bar{\rm C}+\delta {\rm C}<0$ that must liquidate assets first from $\bar\IA$  and when these assets are depleted,  from $\bar{\rm A}$. Selling interbank assets is the same thing as recalling interbank loans, and so this action will inflict additional ``secondary'' liquidity shocks to other debtor banks' liabilities: this domino-like effect is the core of the liquidity cascade mechanism. An illiquid bank that has sold all of  $\bar\IA$ will be called ``fully illiquid'', and must sell external fixed assets  $\bar{\rm A}$ in order to survive. 

We now focus on  ${\tilde p}^{(n)}_i$, the fraction of  bank $i$'s initial interbank assets that remain unsold after $n$ steps of the funding liquidity cascade, starting with ${\tilde p}^{(0)}_i=1$. Illiquid banks will be those with ${\tilde p}^{(n)}_i<1$ while normal banks have ${\tilde p}^{(n)}_i=1$. 

By following logic that parallels the development of equations \eqref{ENcascadeP}-\eqref{ENcascadeQ}, one sees it is now helpful to define $\Sigma^{(n)}_i$, the {\em liquidity buffer} of bank $i$ after $n$ cascade steps, that measures the resilience of the bank to deposit withdrawals. This takes the initial value $\Sigma^{(0)}_i={\rm C}_i^{(0)}$ and as the cascade progresses $\Sigma^{(0)}_i-\Sigma^{(n)}_i$ represents the cumulative amount of deposit withdrawals that the bank has repaid up to day $n$.  This difference is entirely due to the additional deposit withdrawals needed to repay interbank loans, and thus 
\be\label{ENliquidity2}
\Sigma^{(n)}_i=\Sigma^{(0)}_i-\sum_j\Omega^{(0)}_{ij}(1-{\tilde p}^{(n-1)}_j)\ , 
\ee
where the sum is over creditor banks of $i$.
The fraction of $\IA^{(0)}_i$ not sold by day $n$ must be
\be
{\tilde p}^{(n)}_i=\One({\IA}^{(0)}_i>0)\ h( \Sigma^{(n-1)}_i/{\IA}^{(0)}_i)
\label{ENliquidity}\ee
where $h(x)$ is the threshold function \eqref{threshold}.
Finally, the fraction $\tilde q^{(n)}_i$ of bank $i$'s external fixed asset ${\rm A}^{(0)}_i$ not sold by day $n$ must be
\be\label{ENliquidityq} \tilde q^{(n)}_i=\One({{\rm A}}^{(0)}_i>0)\ h\bigl(( \Sigma^{(n-1)}_i+{\IA}^{(0)}_i)/{{\rm A}}^{(0)}_i\bigr)\ .
\ee

The three equations \eqref{ENliquidity2}-\eqref{ENliquidityq} defining the GL cascade model are fully consistent with Assumptions GL. Moreover, comparison with \eqref{ENcascadeP}-\eqref{ENcascadeQ} reveals that the GL cascade model is precisely equivalent to the  EN cascade model described in Section 2.1, with the role of assets/liabilities, equity/cash and liquidity/default buffers simultaneously interchanged: ${\rm A}\leftrightarrow{\ED}, {\IA}\leftrightarrow{\ID},{\rm C}\leftrightarrow{\EE}$,  $\Delta\leftrightarrow{\Sigma}$. It follows that the GL cascade model admits an autonomous stock-flow consistent formulation that is dual to the solvency step $\CS$ described in Section 2.1.1. Each {\em liquidity step} $\CL$ can be understood as an attempt by the system to resolve all banks with a negative cash balance. Just as the solvency step can be viewed as arising under the supervision of a system regulator that dictates the restructuring of insolvent banks, the liquidity cascade arises when the system regulator also determines how illiquid banks should be resolved. As clarified in Section 3, we will interpret negative cash balances as overdrafts drawn from a ``Lender of Last Resort" (LLR) in the sense of \citep{Bagehot1873}. This is a second role of the system regulator in addition to acting as ``bank resolver'' as in Section 2.1.1. Each  {\em liquidity step} $\CL$ maps a collection $[\CB,\Omega]$ of balance sheets and exposures to a new collection $[\CB',\Omega']$ as follows
 \begin{enumerate}
\item[L1] For every illiquid bank $i$  with $-\IA_i\le C_i<0$, a fraction $1-\ \dot\tilde{\! \!p}_i$ of the interbank assets of $i$  are sold, where $\ \dot\tilde{\! \!p}_i=\frac{\IA_i+C_i}{\IA_i}<1$.  If  $C_i<-\IA_i$, then $\ \dot\tilde{\! \!p}_i=0$ and a fraction $1-\ \dot\tilde{\! \!q}_i$ of external assets of $i$  are sold, where $\ \dot\tilde{\! \!q}_i=\frac{{\rm A}_i+\IA_i+C_i}{{\rm A}_i}<1$.
  \item[L2] The unsold (or rolled-over) interbank exposures held by illiquid banks $i$  are valued at $\IA'_i=
  \ \dot\tilde{\! \!p}_i\IA_i, \Omega'_{ji}=\ \dot\tilde{\! \!p}_i\Omega_{ji}$ and the unsold external assets are  valued at ${\rm A}^{F'}_i=\ \dot\tilde{\! \!q}_i {\rm A}_i$.  Based on these asset sales, each illiquid bank $i$  has now raised cash $(1-\ \dot\tilde{\! \!q}_i) {\rm A}_i+(1-\ \dot\tilde{\! \!p}_i)\IA_i=(C_i)^-$ which is  precisely enough to pay the LLR overdraft. 
    \item[L3] However, interbank debt called by illiquid banks must be repaid: All banks $j$  repay $\sum_i(1-\ \dot\tilde{\! \!p}_i) \Omega_{ji}$ of called debt of illiquid banks, leaving the value $\ID'_j=\sum_i \Omega'_{ji}$ of remaining interbank debt. These debt payments lead to new values of liquidity buffers:
  \be C'_j=(C_j)^+-\sum_i(1-\ \dot\tilde{\! \!p}_i) \Omega_{ji}\ .
  \ee
\end{enumerate}

\medskip\noindent{\bf Modifications and Extensions:\ }
\begin{enumerate}
\item The model of \citep{SHLee13} supposes that  instead of GL1, deposit withdrawals are paid by selling liquid external assets (or ``cash'') and interbank assets in equal proportion until depleted, and thereafter liquidating fixed external assets. One can verify that this model is a variation of the above GL framework obtained by introducing a fictitious bank with $i=0$ that borrows $\Omega_{0j}^{(0)}$  from each bank $j$, and by taking $\bar{\rm C}_i=0$ for all $i$. This has the effect of making all the banks initially ``illiquid'' since the initial liquidity buffers are $C^{(0)}_i=\delta \ED_i\le 0$. 
\item The model of  \citep{GaiHalKap11} aims to describe {\em liquidity freezes} observed in past crises and is equivalent to a variation of the GL cascade model that replaces the function $h(x)$ in \eqref{ENliquidity} by $h^\lambda(x)=\lambda\One(x>0)+(1-\lambda)$ for a parameter $\lambda\in[0,1]$. 
  \item One may also take the interbank assets in $\Lambda$ classes ordered by degree of liquidity, with class $\ell=1$ the most liquid class. For example, one might assume that $\ell$ labels interbank loans of different maturity (including the external loans as lending to bank $i=0$), which the banks liquidate in increasing order.  
  \item Finally, one must expect that illiquid banks incur additional expenses or liquidation penalties, for example punitive interest rates charged by the LLR or penalties for early withdrawal of interbank loans. Such cash penalties  further erode both the bank's cash buffer and equity buffer.
%
\end{enumerate}


%
%
%
%
%
%

\section{Solvency and Liquidity Cascade Model}
\label{sec:dual}

The stock-flow consistent forms of the EN solvency cascade model of Section 2.1.1 and the GL cascade model of Section 3.1 can be  intertwined in a natural way to give a financial network cascade model unifying the two channels of direct contagion (i) and (ii) discussed in the Introduction. The resultant Solvency and Liquidity (SL) cascade model which we now describe rests on simple yet financially reasonable assumptions. In fact, we propose it as the simplest possible financially justifiable cascade model that like the model of \cite{HurdCelMelSha16} unifies solvency and liquidity.  
 
As in the previous models, the post-trigger initial balance sheet of bank $i$ has the form $B^{(0)}_i=[\IA^{(0)}_i,{\rm A}^{(0)}_i,{\rm C}^{(0)}_i,\ID^{(0)}_i, \ED^{(0)}_i,\EE^{(0)}_i]$. Following the triggering event, which might be a combination of major external deposit withdrawals or external fixed asset price declines, the system at this moment is assumed to be {\em impaired}, which means some of the banks have either ${\rm C}^{(0)}_i<0$ (illiquid) or ${\rm E}^{(0)}_i<0$ (insolvent).  

The resultant crisis cascade is interpreted mechanistically as a specification of how the entire system, guided by a {\em system regulator},  iteratively rebalances its balance sheets to resolve bank insolvencies and repay bank overdrafts, aiming each day to restore compliance with all accounting constraints. It does so by applying two steps alternately. During the first step $\CS$, called the {\it restructuring} or {\em solvency} step, the insolvent banks in the system undergo a bail-in restructuring of their debt in an attempt to restore their negative equity to zero. Subsequently, however, the systemic effect of interbank debt restructuring causes a further negative impact on the network likely to cause additional bank insolvencies. In the second step $\CL$, called the {\it asset liquidation} or {\em liquidity} step, illiquid banks attempt to restore their negative cash position to zero by recalling loans, or as we prefer to say, by selling interbank and fixed assets. Subsequently, however, the systemic effect of interbank asset liquidation is a further negative impact on the network in the form of additional interbank funding illiquidity. 

The SL cascade model is based on Assumptions A1-A4 and the following more refined versions of assumptions EN1-EN2 and GL1-GL2.  
\begin{enumerate}
  \item[SL1] There is a {\it system regulator} that combines two functions\begin{enumerate}
  \item It acts as a lender of last resort (LLR)\footnote{Think of this as Bagehot's concept \citep{RochVive04}.}  that lends to illiquid banks (with ${\rm C}<0$) to allow them to make essential payments to other banks. Such loans are without interest, and must be repaid within one period. 
  \item It enforces the minimal restructuring of insolvent banks' debt (with $\EE<0$) to return their equity to zero,   through ``bailing in'' of creditors that take a loss on their insolvent exposures.
\end{enumerate}
\item[SL2] Interbank lending is overnight, and illiquid banks always make required payments on maturing interbank exposures with cash, or if illiquid, by borrowing from the LLR. 
  \item[SL3] External debt is senior to interbank debt, which implies insolvent banks pay all their external debt  before paying any interbank debt. External assets are less liquid than interbank assets, which implies illiquid banks  sell their interbank assets before selling any external assets. 
   \item[SL4]  There no direct losses, costs or penalties due to bankruptcy or illiquidity. 
  \item[SL5] There is no market price impact when interbank and external assets are sold. External depositors are fully insured and have no incentive to withdraw funding from failing banks. 
\end{enumerate}

\medskip
\noindent{\bf Remark:\ } If necessary, external debt of equal seniority to interbank debt can be included as debt from a ``fictitious'' bank $i=0$ that is never insolvent or illiquid. Similarly, external assets of equal liquidity to interbank assets can be included as loans to this fictitious bank $i=0$.

\medskip
We now specify how balance sheets and exposures on day $n-1$,  $B^{(n-1)}_i=[\IA^{(n-1)}_i,$\\${\rm A}^{(n-1)}_i,{\rm C}^{(n-1)}_i,\ID^{(n-1)}_i, \ED^{(n-1)}_i,\EE^{(n-1)}_i]$ and  $\Omega^{(n-1)}_{ij}$, are updated to day $n$. This SL cascade mapping $\CC$ is a composition $\CC=\CL\circ\CS$ as the system regulator applies first the solvency step $\CS$ as specified by S1-S3, and then the liquidity step $\CL$ as specified by L1-L3.

\bigskip
\noindent{\bf SL Cascade Mapping:\ } The mapping $\CC=\CL\circ\CS$ from day $n-1$ to day $n$ acts on the data $[\Omega^{(n-1)}_{ij}, p^{(n-1)}_i,\tilde p^{(n-1)}_i, \EE^{(n-1)}_i,{\rm C}^{(n-1)}_i] $ as follows
\begin{enumerate}
  \item The exposures at step $n-1$ are $\Omega^{(n-1)}_{ij}=p^{(n-1)}_i\tilde p^{(n-1)}_j\Omega^{(0)}_{ij}$ and hence 
  \be
  \label{ZX}\ID^{(n-1)}_i=p^{(n-1)}_i\sum_j\tilde p^{(n-1)}_j\Omega^{(0)}_{ij}, \IA^{(n-1)}_i=\tilde p^{(n-1)}_i\sum_jp^{(n-1)}_j\Omega^{(0)}_{ji}\ .
  \ee  
  \item The solvency step $\CS$ produces
 \beq
   \label{pupdate}p^{(n)}_i&=&p^{(n-1)}_i\dot p^{(n)}_i,\ \dot p^{(n)}_i=\One(\ID^{(n-1)}_i>0) h(\EE^{(n-1)}_i/\ID^{(n-1)}_i)\\
   \label{Deltaupdate} {\rm E}^{(n)}_j&=&(\EE^{(n-1)}_j)^+-\sum_i(1- \dot p^{(n)}_i) \Omega^{(n-1)}_{ij}\ 
   \eeq
and intermediate values for interbank assets\be
    \label{Zupdate} \tilde \IA^{(n)}_i=\sum_j\dot p^{(n)}_j\Omega^{(n-1)}_{ij}\ .
    \ee   \item The liquidation step $\CL$ produces
\beq
   \label{tpupdate}\tilde p^{(n)}_i&=&\tilde p^{(n-1)}_i \ \dot\tilde{\! \!p}^{(n)}_i,\quad \ \dot\tilde{\! \!p}^{(n)}_i=\One(\tilde\IA^{(n)}_i>0)h({\rm C}^{(n-1)}_i/\tilde\IA^{(n)}_i)\\ 
\label{Sigmaupdate} {\rm C}^{(n)}_j&=&({\rm C}^{(n-1)}_j)^+-\dot p^{(n)}_j\sum_i(1-  \ \dot\tilde{\! \!p}^{(n)}_i) \Omega^{(n-1)}_{ji}\ .
\eeq
\item One also reconstructs the solvency and liquidity buffers:
\beq
\label{Dn}\Delta^{(n)}_i&=&{\rm E}^{(n)}_i-\sum_{m=0}^{n-1}({\rm E}^{(m)}_i)^-\\
\label{Sn}\Sigma^{(n)}_i&=&{\rm C}^{(n)}_i-\sum_{m=0}^{n-1}({\rm C}^{(m)}_i)^-\ .
\eeq
\item The external fixed assets and debts
${\rm A}_i^{(n)}=\tilde q^{(n)}_i{\rm A}_i^{(0)},\ \ED^{(n)}_i=q^{(n)}_i\ED^{(0)}_i$ are determined in terms of supplementary quantities $q^{(n-1)}_i, \tilde q^{(n-1)}_i$:
\beq
\label{firesale1}q^{(n)}_i&=&q^{(n-1)}_i\ \ \dot q^{(n)}_i\\
\label{firesale2} \dot q^{(n)}_i&=&\One(\ED^{(n-1)}_i>0)  h\bigl((\EE^{(n-1)}_i+\ID^{(n-1)}_i)/\ED^{(n-1)}_i\bigr),\\
\label{firesale3}\tilde q^{(n)}_i&=&\tilde q^{(n-1)}_i\ \dot\tilde{\! \!q}^{(n)}_i\\
\ \dot\tilde{\! \!q}^{(n)}_i&=&
\ \One({\rm A}^{(n-1)}_i>0) h\bigl(({\rm C}^{(n-1)}_i+\tilde \IA^{(n)}_i)/{\rm A}^{(n-1)}_i\bigr)\ .
\label{firesale4}\eeq

\end{enumerate}

It is important to note that Steps 1-3 of the SL cascade mapping form an autonomous system of equations for a reduced set of variables 
$[p^{(n-1)}_i,\tilde p^{(n-1)}_i, \EE^{(n-1)}_i,$\\${\rm C}^{(n-1)}_i] $, depending also on the initial exposure matrix $\Omega^{(0)}_{ij} $ but in particular do not depend on ${\rm A}^{(0)},\ED^{(0)}$. Of course, Step 5 gives the impact of the cascade on the external assets and debt, and does  depend on the initial data ${\rm A},\ED$.

Given that the SL mapping is essentially an AL symmetric version of the Eisenberg-Noe model, the following result is not surprising. However, the proof requires an additional step. 

\begin{theorem}\label{SLtheorem} The limit $\lim_{n\to\infty}(\CL\circ\CS)^{n}(\CB^{(0)})$ starting from the initial state $\CB^{(0)}$  with exposures $\Omega^{(0)}$ and ${\rm p}^{(0)}=\tilde{\rm  p}^{(0)}=\One$ is the maximal fixed point $\hat{\CB}$ of
\[\CB=(\CL\circ\CS)(\CB)
\] and has the following properties:
\begin{enumerate}
\item $\hat{\CB}=\CS(\hat{\CB})=\CL(\hat{\CB})$.
  \item $\hat{\rm E}_j\ge 0$. The bank is insolvent if and only if $\hat{\rm E}_j=0$.
  \item $\hat{\rm C}_j\ge 0$. The bank is illiquid if and only if $\hat{\rm C}_j=0$.
 \end{enumerate}
\end{theorem}

Thus, the cascade is monotonically increasing in severity, and ends in an equilibrium state where all insolvent banks have exactly zero equity, and all illiquid banks have exactly zero cash assets and owe zero to the LLR. One can also see that if $p^*_i=0$, then the bank becomes fully insolvent (meaning no interbank debt can be repaid). Similarly, if $\tilde p^*_i=0$, then the bank ends up fully illiquid (meaning all interbank assets have been sold).

\bigskip\noindent{\bf Proof:\ } We wish to apply the Knaster-Tarski Fixed Point Theorem as done for the Eisenberg-Noe result (Theorem 1) but the mapping $\CC=\CL\circ\CS$ defined by \eqref{ZX}-\eqref{Sigmaupdate} is not monotonically decreasing in the variables $[p^{(n)}_i,\tilde p^{(n)}_i, {\rm E}^{(n)}_i,{\rm C}^{(n)}_i]_{i\in[N]}$. However, the following mapping $\tilde\CC$ on the variables  $[p^{(n-1)}_i,p^{(n)}_i,\tilde p^{(n-1)}_i, \tilde p^{(n)}_i, ({\rm E}^{(n-1)}_i)^+,$\\$({\rm C}^{(n-1)}_i)^+]$ is monotonic and bounded on the set $L$ of points $\ell:=[x_i,y_i,\tilde x_i,\tilde y_i, {\rm E}_i,$\\${\rm C}_i]_{i\in[N]}\in\BBR^{6N}$ satisfying the inequalities $0\le y_i\le x_i\le  1$, $0\le \tilde y_i\le \tilde x_i\le 1$, $0\le {\rm E}_i\le ({\rm E}^{(0)}_i)^+$, $0\le {\rm C}_i\le ({\rm C}^{(0)}_i)^+$ for each $i\in[N]$
\begin{enumerate}
  \item Compute ${\rm E}^{(n)}_i$ from \eqref{Deltaupdate}, and note that $({\rm E}^{(n)}_i)^+\le ({\rm E}^{(n-1)}_i)^+$;
   \item Compute ${\rm C}^{(n)}_i$ from \eqref{Sigmaupdate}, and note that $({\rm C}^{(n)}_i)^+\le ({\rm C}^{(n-1)}_i)^+$;
     \item Compute $p^{(n+1)}_i$ from \eqref{pupdate} with $\ID^{(n)}_i$ computed from \eqref{ZX}, and note that $p^{(n+1)}_i\le p^{(n)}_i $;
    \item Compute $\tilde p^{(n+1)}_i$ from \eqref{tpupdate} with $\tilde\IA^{(n+1)}_i$ computed from \eqref{Zupdate}, and note that $\tilde p^{(n+1)}_i\le\tilde p^{(n)}_i $.
\end{enumerate}
Moreover,  $\tilde\CC$ maps the set $L$ onto itself. Since $L$ is a complete lattice, the existence of fixed points of $\tilde \CC$ is guaranteed by the Knaster-Tarski Theorem. 

Since $\tilde\CC$ is also a continuous mapping, the iteration scheme starting from the maximal initial state $\ell^{(0)}=[\One,\One,\One,\One,({\rm E}^{(0)})^+,({\rm C}^{(0)})^+]$  with exposures $\Omega^{(0)}$  must converge to the maximal fixed point $\hat{\ell}$ of the mapping $\tilde\CC$. One can easily verify that $\hat\ell$ has the form $[\hat{\rm p},\hat{\rm p},\hat{\tilde{\rm p}},\hat{\tilde{\rm p}},(\hat{\rm E})^+,(\hat {\rm C})^+]$ where $(\hat{\rm E})^+=\hat{\rm E}$ and $(\hat{\rm C})^+=\hat{\rm C}$. In other words, Properties 2 and 3 of the Theorem hold. Clearly also $\hat\ell$ determines the maximal fixed point $\hat\CB$ of $\CC$, which necessarily satisfies Property 1.

\noindent{\bf The clearing condition:\ } It appears one cannot characterize the fixed points of $\CC$ in terms of a lower dimensional mapping analogous to the $N$ dimensional EN clearing condition ${\rm p}={\rm F}({\rm p})$. For example, unlike \eqref{ENcascadeD} from that single cascade model, the information of the vectors ${\rm p}^{(n)},\tilde {\rm p}^{(n)}$ for fixed $n$ is not sufficient to determine the buffers $\Delta^{(n)},\Sigma^{(n)}$.

%
%
%
%
\section{Extended SL Cascade Model}
Bank insolvencies and illiquidity both induce shocks that are mediated directly through the interbank exposures. Such {\em direct feedback} effects are {\it internal} to the financial network. A complete picture of systemic risk must also consider  {\em indirect feedback} between the financial system and the external economy, represented in our framework by the external fixed assets ${\rm A}$ and external debt $\ED$. The SL cascade model of Section 3 cannot be taken seriously as a model of systemic risk, because Assumption SL5 rules out indirect feedback effects. The {\em extended solvency and liquidity} (ESL) cascade model we now introduce builds in stylized indirect feedback effects that overcome Assumption SL5.

The literature on indirect contagion separates into asset side contagion and liability side contagion. External asset contagion is often called the {\it asset fire sale} effect, and has been understood as a primary component in historical financial crises (see \citep{ShleVish92,CifuFerrShin05} and references cited therein). A  quantitative discussion of the role of securitization and the fire sales that took hold in these products during the 2007-08 crisis can be found in \citep{Shin09}. On the other hand, external liability contagion,  which has also been called {\it rollover risk} or {\it bank panic risk}, is also thought to be a significant component of crises (see \citep{CaloGort91,RochVive04,XuewenLiu16} and references cited therein). The ESL cascade model views external asset and liability contagion as two sides of the same coin, and models them accordingly,  providing a minimal description of both phenomena that can anchor more complex explanations. A recent paper \citep{AldaFaia16} provides an alternative framework that similarly combines asset side contagion and liability side contagion in a network model for optimizing banks subject to regulatory constraints.

\subsection{Asset fire sales} Let us follow \citep{CifuFerrShin05} in supposing for simplicity that banks hold shares of a common fixed external asset. This 100\% asset correlation assumption can be viewed as a worst-case assessment of the diversification strategies adopted by banks, and leads to the strongest possible fire sale effect. An extended network model with multiple asset classes has been introduced by \citep{CacShrMooFar14}.   Without loss of generality, this common asset is assumed to have an initial share price $\Pi^{(0)}=1$ of which each bank $i$  initially holds ${\rm A}^{(0)}_i$ units.  At step $n$ of the crisis,  the liquidation step $\CL$ implies that some illiquid banks may sell fixed assets: based on equation \eqref{firesale3}, the number of units held by bank $i$   will be reduced from  $\tilde q^{(n-1)}_i {\rm A}^{(0)}_i$ to $\tilde q^{(n)}_i {\rm A}^{(0)}_i$.  Other banks may also reduce their appetite for purchasing additional units of the external asset.  Since financial institutions are large traders, such asset sales combined with reduced demand may have significant {\it market impact}, meaning they cause a permanent downward shift in the underlying asset price $\Pi$. Such  asset price changes will lead to declines in external asset values ${\rm A}^{(n)}_i=\Pi^{(n)}\tilde q^{(n)}_i {\rm A}^{(0)}_i$ of every bank in the network that are one-to-one with declines in their equity buffers.  

We assume the final asset price paid by the external market at the end of any period is given by a pricing function $P(s-d)$ of $s-d$, the excess supply over demand for the asset  by the banking system over this period. That is, $s-d$ is the difference between the total amount sold $s$ and the change in system demand $d$,  over the period. We also assume that the system demand is linear in the liquid assets in the system, so that  $d=\mu \ell+\mu'\ell'$, where $\ell$ and $\ell'$ are respectively the total amount of liquid interbank and cash assets added by the system and the constants $\mu, \mu'\in[0,1]$ represent the capacity of these liquid funds to purchase the fixed asset. Note that the available liquidity of the interbank and cash market, and hence $d$, all decrease during the crisis.  The pricing function $P$ should be monotonically decreasing and convex in $s-d$ and thus in the following we adopt the choice of \citep{CifuFerrShin05}
\be \label{inversedemand} P(s-d)=\Pi^{(0)}e^{-\alpha (s-d)}
\ee
where $\Pi^{(0)}=1$ is the price at the beginning of the period. The  positive constant $\alpha^{-1}$  represents the depth of the demand for the fixed asset by the external market: a heuristic rule that the price is reduced by 50\% when the system sells 100\% of its fixed assets leads to the value $\alpha=\log 2/\sum_i {\rm A}_i$. 

The $\CL^A$ step of the ESL cascade model is a modification of $\CL$ that accounts for the impact of the decline in the asset price $\Pi$ on the equity of banks. In terms of the quantities described in Section \ref{sec:dual}, \begin{enumerate}
  \item The total amount of external asset  sold $s$, the total amount of interbank assets sold $\ell$,   and the total amount  of cash assets  lost $\ell'$ change  recursively as banks become illiquid:  
  \beq \label{excesssupply} s^{(n)}&=&s^{(n-1)}+\sum_i (1- \ \dot\tilde{\! \!q}^{(n)}_i)\tilde q^{(n-1)}_i {\rm A}^{(0)}_i,\ s^{(0)}=0\\
  \label{excessdemandZ}  \ell^{(n)}&=&\sum_i ({\rm Z}^{(n)}_i-{\rm Z}^{(0)}_i)\\
  \label{excessdemandC}  \ell^{'(n)}&=&\sum_i ({\rm C}^{(n)}_i-{\rm C}^{(0)}_i)\ .\ \eeq
\item  The asset price is determined recursively by
\beq\label{priceeqn} \Pi^{(n)}&=&\Pi^{(n-1)}\\
&&\hspace{-1in}\times\ \exp\left[-\sum_i\left(\alpha (1-\dot\tilde{\! \!q}^{(n)}_i)\tilde q^{(n-1)}_i {\rm A}^{(0)}_i+\beta(\IA^{(n-1)}_i-\IA^{(n)}_i)+\beta'  ({\rm C}^{(n-1)}_i)-{\rm C}^{(n)}_i)\right)\right]\nonumber \eeq
with $ \Pi^{(0)}=1$ and $\beta=\alpha\mu,\beta'=\alpha\mu'$. 
\item Since changes in the mark-to-market value of fixed assets are one-to-one with changes in the equity buffer there is an additional change in banks' equity:
 \be\label{Deltaupdate2} \EE^{(n)}_i= \tilde\EE^{(n)}_i-(\Pi^{(n-1)}-\Pi^{(n)})\tilde q^{(n)}_i {\rm A}^{(0)}_i\ ,\ee 
where $\tilde\EE^{(n)}_i$ denotes the intermediate equity buffer at the end of the solvency step.\end{enumerate}

In summary, inclusion of the fire sale effect into the SL framework amounts to an additional impact to the solvency buffers through equation \eqref{Deltaupdate2}, over and above the impact due to devaluation of interbank assets due to defaults. 

%
%

\subsection{Bank panics and external funding shocks} A {\em bank panic} can be said to occur when the holders of external debt (which we call {\em depositors}) collectively lose confidence in the banking system, and therefore make a significant withdrawal of funding to the system. Since this second channel of indirect contagion has been little studied in the economics literature, we find it helpful to model it as the AL symmetric analogue of the single asset fire sale mechanism described above.  Therefore, in the following, the collection of depositors  is viewed as a {\em single pool} whose confidence in the financial system as a whole is reduced by debt restructuring and drops in equity of any one bank. On day $n$, the depositor pool holds $\tilde \Pi^{(n)} \ED^{(0)}_i$ units  of the external debt of bank $i$, with a unit price $q^{(n)}_i$ that evolves through equation \eqref{firesale1} of the solvency step $\CS$, giving a total value 
\be\label{EDN}
\ED^{(n)}_i=\tilde \Pi^{(n)}q^{(n)}_i \ED^{(0)}_i \ . 
\ee
The single pool assumption has the meaning that debt restructuring or a decrease in equity of any one bank through the solvency step $\CS$  has the same proportional effect on the fractional amount $\tilde \Pi^{(n)}$ of all banks' external debt. Note that this is a strong assumption that implies for example that nervous depositors are unable to identify and run on particularly weak banks.

We have seen in Section 3 that a deposit withdrawal is the AL symmetric dual of an asset price drop: both lead to a one-to-one impact on one buffer while leaving the other buffer unchanged. Bankers directly control the number of units of fixed assets they hold by trading, but cannot directly control the asset price. Analogously,  bank $i$ cannot directly control the number of units of debt since it accepts all new deposits and honours all deposit withdrawals, but the unit price of its debt $q^{(n)}_i$ is reduced through restructuring. 

We therefore suppose that the total fractional amount of external debt $\tilde \Pi$ is negatively related to the total amount of restructured external debt $\tilde s$, the total interbank debt lost $\tilde d$, and the total banking system equity lost $\tilde e$.
These dependencies are determined through a deposit impact function $\tilde \Pi=\tilde  P(\tilde s,\tilde d,\tilde e)$ for the fraction of the initial system-wide aggregated external debt that is not yet withdrawn. In parallel with \eqref{inversedemand} we suppose a linear relation between $\tilde s,\tilde d,\tilde e$ and $\log \tilde \Pi$:
\be\label{withdrawal}  \tilde  P(\tilde s,\tilde d,\tilde e)=e^{-\tilde\alpha \tilde s-\tilde \beta\tilde d-\tilde \beta' \tilde e}\ee
where $\tilde\alpha, \tilde \beta, \tilde \beta'$ are {\em confidence weakening} parameters. A formula similar to \eqref{withdrawal}, given in \citep{AldaFaia16}, can be microfounded using the theory of global games \citep{MorrShin03}. 

The $\CS^D$ step of the ESL cascade model is a modification of the $\CS$ step that incorporates the bank panic contagion mechanism by accounting for the impact on cash buffers of all banks due to additional withdrawals by depositors. At the end of the $\CS^D$ step on day $n$ of the crisis each bank $i$  has external debt valued by \eqref{EDN}.  \begin{enumerate}
  \item The total amount of restructured external debt $\tilde s$, the total interbank debt lost $\tilde d$ and the total bank equity lost $\tilde e$ all grow as banks become increasingly insolvent:  
  \beq
  \label{panic1}  \tilde s^{(n)}&=&\tilde s^{(n-1)}+\sum_i (1-\dot q^{(n)}_i)q^{(n-1)}_i\ED^{(0)}_i\ ,\  \tilde s^{(0)}=0 \\ 
   \label{panic2}  \tilde d^{(n)}&=&\sum_i( \ID^{(0)}_i- \ID^{(n)}_i)\ \\
 \label{panic3}   \tilde e^{(n)}&=& \sum_i( \EE^{(0)}_i- \EE^{(n)}_i)\ .
   \eeq
\item 
The total fraction of external deposits not yet withdrawn is
\beq\label{DebtImpact} \tilde\Pi^{(n)}&=&\tilde\Pi^{(n-1)} \\&&\hspace{-.75in}\times \exp\Bigl(-\sum_i \bigl(\tilde\alpha(1-\dot q_i^{(n)})q^{(n-1)}_i\ED^{(0)}_i+\tilde\beta( \ID^{(n-1)}_i- \ID^{(n)}_i)+\tilde\beta'( \EE^{(n-1)}_i- \EE^{(n)}_i)\bigr)\Bigr)\nonumber\eeq 
while the fraction of external deposits withdrawn on day $n$ is $\tilde\Pi^{(n-1)}-\tilde\Pi^{(n)}$.
\item The cash buffers at the end of the $\CS^D$ step, after accounting for the external debt withdrawals that must be repaid, are
\be\label{panic5} \tilde{\rm C}^{(n)}_i= {\rm C}^{(n-1)}_i-(\tilde\Pi^{(n-1)} -\tilde\Pi^{(n)})q^{(n)}_i \ED^{(0)}_i\ .\ee
 \end{enumerate}
 
From equation \eqref{panic5} one sees clearly how external deposit withdrawals, caused by the dwindling confidence of the external lenders in the solvency of the financial system, impact the cash positions of banks. The extent to which this effect can influence the development of a financial crisis is thus ready to be understood within the ESL framework. 
%
%
%
%

\subsection{ESL Cascade Mapping} We have seen that the asset fire sale contagion channel can be incorporated as a modification $\CL^A$ of the liquidity step $\CL$. The bank panic contagion channel is incorporated as a modification $\CS^D$ of the solvency step $\CS$. The complete ESL cascade model amounts to iterating the composition $\CL^A\circ \CS^D$, starting with the post-trigger balance sheets $\CB^{(0)}$ and exposures  $\Omega^{(0)}$:
\be
\label{ESLmapping}
(\CB^{(n)},\Omega^{(n)})=\CL^A\bigl(\CS^D(\CB^{(n-1)},\Omega^{(n-1)})\bigr), \quad n\ge 1\ .
\ee
 Comparison of equations \eqref{panic1}-\eqref{panic5} with equations \eqref{excesssupply}-\eqref{Deltaupdate2} shows clearly how our assumption of AL duality between the $\CL^A$ and $\CS^D$ mechanisms preserves the symmetry of the ESL cascade model.

It is not difficult to extend the proof of Theorem 3 to the ESL Cascade Mapping, leading to the obvious conclusions about the convergence of the cascade iteration scheme to a maximal fixed point.

%
%

\begin{figure}
  \centering

 \def\svgwidth{\columnwidth}{\scriptsize
  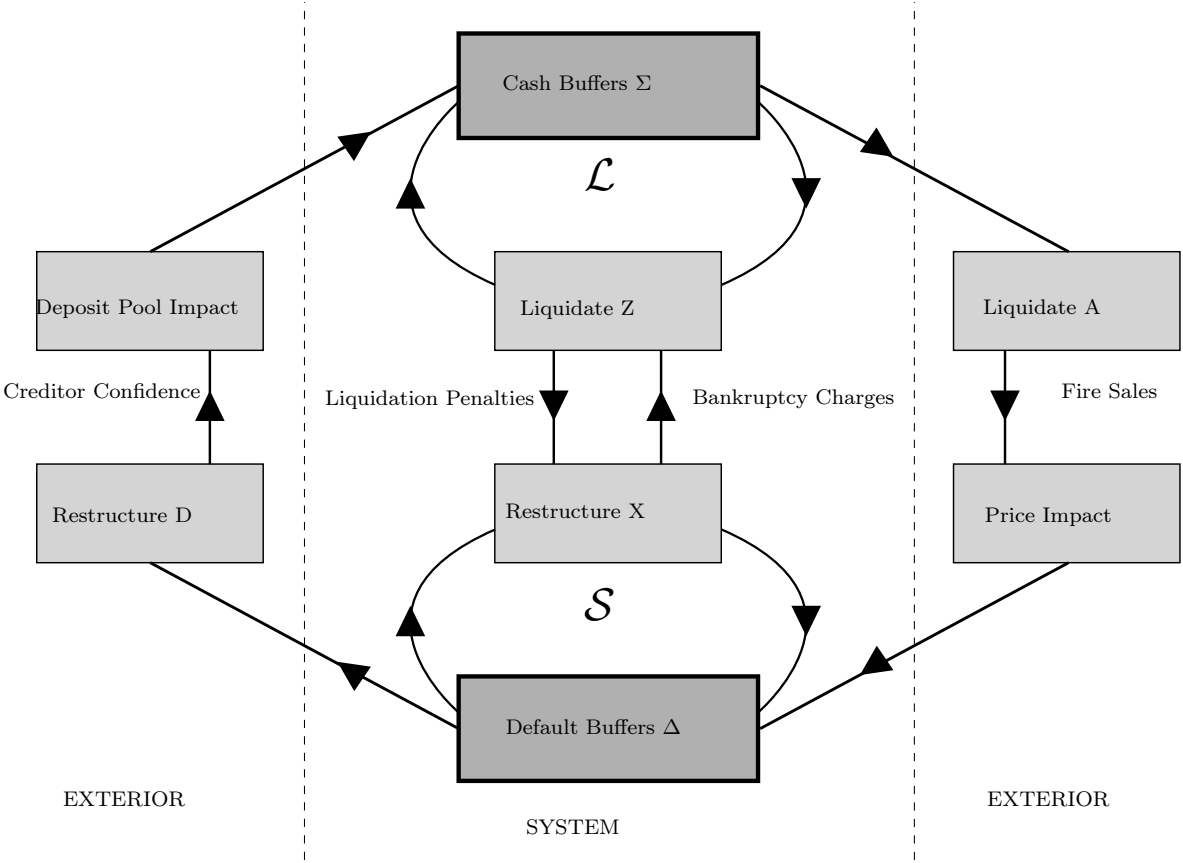}
  \caption{A schematic representation of the ESL cascade model, showing how the two direct contagion channels $\CL$ and $\CS$ intertwine with the two indirect contagion channels, fire sales and bank panics. Two important effects not in the basic ESL cascade model described in Section 4, namely liquidation penalties and bankruptcy charges, are also included in this diagram for completeness. }
  \label{crisiscycle}
   \end{figure}

Figure \ref{crisiscycle} shows schematically where the various types of shocks incorporated in the ESL cascade model are created and have impact. We emphasize that the geometric symmetries of the figure correspond to consequences of the AL symmetric modelling assumptions we have made, and are realized in symmetries of dynamics within the ESL cascade mapping. 

Finally, we note that the missing arrows in Figure 1 corresponding to various types of liquidation and bankruptcy penalties, are easily incorporated in the ESL cascade models by including additional negative terms in equations \eqref{Deltaupdate2} and \eqref{panic5}.  

\section{Discussion}
\label{sec:conc}

The most important insights, and potential criticisms, of this paper stem from the imposition of stock-flow consistency on systemic risk models. In hindsight, Theorem 2 states a somewhat obvious fact, not previously noticed,  about the original model of \citep{EiseNoe01}. However, viewing the EN clearing condition as resulting from an iterated stock-flow consistent ``solvency step'' brings in the interpretation of an ideal system regulator that enforces an orderly ``bail-in'' resolution mechanism for insolvent banks. Since the reality of bank resolution is much messier and more damaging than how we have modelled it here, one has a strong reason for being skeptical of the EN model as a systemic risk model of insolvency. Analogously, in the Generalized Liquidity cascade model one finds the need for the regulator to act as a lender of last resort to ensure functioning of the interbank lending market during a crisis. In this case, however, the iterated liquidation step is a reasonable picture of how banks manage deposit withdrawals. 

Just as stock-flow consistency leads to a clear picture of insolvency and illiquidity viewed separately, the Solvency-Liquidity (SL) cascade model provides a plausible characterization of how these two cascades intertwine. However, without the two channels of indirect contagion, namely asset fire sales and bank panics, insolvency and illiquidity are nearly independent in the sense that the cash buffer is unaffected by the solvency step and the equity buffer is unaffected by the liquidation step.  Since such independence is implausible in reality, only the Extended SL cascade model that incorporates two channels of indirect contagion meets our criteria for a comprehensive cascade model of systemic risk that fully addresses spillovers between insolvency and illiquidity. In fact, one can argue that a model that does not at least address all the contagious flows shown in Figure 1 cannot be taken as a comprehensive systemic risk model. 

The particular way indirect contagion is included in the ESL cascade model makes additional simplifying assumptions whose implications are important to understand. First, the asset fire sale mechanism is based on banks holding a single common external fixed asset. In other words, their investment portfolios are assumed to be 100\% correlated. The reality is that in most countries,  bank portfolios are indeed highly, but not perfectly correlated. Since we expect systemic risk to be increasing with portfolio homogeneity, the common asset assumption is a prudent (i.e.  pessimistic) assumption. It is also parsimonious: accounting for multiple asset types along the lines of \citep{CacShrMooFar14} will add a considerable layer of complexity to the ESL cascade model. Analogously, we have treated external depositors as a single homogeneous pool that relies on system-wide indicators of systemic risk and is not able to spot individual weak banks. Consequently, at times of crisis they withdraw funding from the system en masse, causing bank panics but not bank runs. Again, a more realistic model will follow \citep{RochVive04} and disaggregate the deposit pool, and enable depositors of weak banks to run more strongly than depositors of strong banks. However, we expect systemic risk to be increasing in depositor homogeneity, so our single pool assumption is again both prudent and parsimonious. 

Another reason to take the ESL cascade model seriously is that it has a rich family of parameters that can be adjusted to focus on the various contagion channels separately or in combination. For example,  setting the parameters $\alpha, \beta, \beta'$ all zero turns off the fire sale effect, allowing one to test the significance of this channel in propagating contagion. Moreover, having the full range of parameters of the ESL cascade model to work with should enable one to mimic most of the existing contagion models identified in the literature review in the introduction, to compare the explicit and hidden assumptions they make, and to test whether or not they omit significant features.

Any comprehensive systemic risk model that accounts for all the flows identified in Figure 1 will be at least as complicated as the ESL cascade model, and will be based on a large number of behavioural assumptions, many of which will likely be ad hoc rather than microfounded. The ESL cascade model relies on Asset-Liability symmetry to reduce the number and complexity of assumptions needed, and appeals to Occam's Razor for its justification. While there are good reasons to be skeptical of AL symmetry in the real world, there are also good reasons not to abandon it as a tool in systemic risk modelling.  There is insight gained in realizing that a mark-to-market decrease in fixed asset valuations is the AL dual to a new debt withdrawal, and that similarly, a new loan added to the books that depletes the cash buffer is the AL counterpart to a mark-to-market increase in the value of debt that depletes the equity buffer. While we may fail to see the AL symmetry at the level of houses (banks), it becomes clearer at the level of the village (the entire network).

Ultimately, the ESL cascade model is intended as a provisional, consistent framework for financial systemic risk comprehensive enough to include all the spillover effects identified by the arrows shown in Figure 1. Exploring some of its implications through simulation experiments based on real-world financial networks will be the subject of a subsequent paper. This model can also be seen as a logical platform upon which successive layers of future improvements can be built.  Questions of which are the most important effects missing from the ESL cascade model, how to model them, and how they change the stability of the system, are a rich source of topics to be investigated in future work. 

\section{Acknowledgements} This project was funded by the Natural Sciences and Engineering Research Council of Canada. The author also gratefully acknowledges the support of the Forschungsinstitut f\"ur Mathematik at ETH Z\"urich and the Centre Interfacultaire Bernoulli at EPF Lausanne,  as well as extensive discussions with Grzegorz Halaj of the European Central Bank and Kartik Anand of Deutsche Bundesbank.

\bibliographystyle{abbrvnat}


\end{document}